\documentclass[aps,prl,preprint,showpacs]{revtex4}
\usepackage{amssymb}
\usepackage{bbm}

\usepackage{amsfonts}
\usepackage{amsmath}
\usepackage{mathrsfs}
\begin{document}

\title{True radiation gauge for gravity}

\author{Xiang-Song Chen$^{1,2,3,}$}
\email{cxs@hust.edu.cn}
\author{Ben-Chao Zhu$^1$}

\affiliation{$^1$Department of Physics, Huazhong
University of Science and Technology, Wuhan 430074, China\\
$^2$ Joint Center for Particle, Nuclear Physics and Cosmology,
Nanjing 210093, China \\
$^3$Kavli Institute for Theoretical Physics China, CAS, Beijing
100190, China}

\date{\today}

\begin{abstract}
Corresponding to the similarity between the Lorentz gauge
$\partial_\mu A^\mu=0$ in electrodynamics and
$g^{\mu\nu}\Gamma^\rho_{\mu\nu}=0$ in gravity, we show that the
counterpart of the radiation gauge $\partial_iA^i=0$ is
$g^{ij}\Gamma^\rho_{ij}=0$, in stead of other forms as discussed
before. Particularly: 1) at least for a weak field,
$g^{ij}\Gamma^\rho_{ij}=0$ fixes the gauge completely and picks out
exactly the two physical components of the gravitational field; 2)
like $A^0$, the non-dynamical components $h_{0\mu}$ are solved
instantaneously; 3) gravitational radiation is generated by the
``transverse'' part of the energy-momentum tensor, similar to the
transverse current $\vec J_\perp$. This ``true'' radiation gauge
$g^{ij}\Gamma^\rho_{ij}=0$ is especially pertinent for studying
gravitational energy, such as the energy flow in gravitational
radiation. It agrees with the transverse-traceless (TT) gauge for a
pure wave, and reveals remarkably how the TT gauge can be adapted in
the presence of source.

\end{abstract}
\maketitle

Gauge invariance is a powerful guidance in building field theories,
but can be a nuisance in actual calculations. Its advantage and
disadvantage are just the two sides of the same coin: The redundant
field components help to write down an elegant Lagrangian, but have
to be got rid of when identifying the real physical degrees of
freedom. In electrodynamics, one uses a four-component vector field
$A^\mu$ to describe a massless photon with two physical
polarizations. The radiation (or Coulomb, transverse) gauge $\vec
\partial \cdot \vec A=0$ can perfectly remove the gauge freedom and
specify the transverse field $\vec A_\perp$ to be the two physical
components which propagate in electromagnetic radiation. In general
relativity, one uses a ten-component symmetric tensor $g_{\mu\nu}$
to describe the gravitational field, but the number of physical
degrees of freedom is still two, thus, naturally, the gauge fixing
problem is much harder. A gravitational ``radiation gauge'' as
satisfactory as $\vec \partial \cdot \vec A=0$ in electrodynamics
should meet the following criteria:
\begin{enumerate}
\item It contains four and only four realizable constraints.

\item It fixes the gauge completely, and picks out exactly the two
physical components of the gravitational field.
\end{enumerate}
So far as we know, such a gauge has not yet been reported. In this
paper, we show that it does exist, at least for linearized gravity.
It differs remarkably from other gauges discussed before, and leads
to interesting implications concerning gravitational radiation and
gravitational energy.

When handling gravitational radiation, the most frequently used
gauge is the harmonic or De Donder gauge
$g^{\mu\nu}\Gamma^\rho_{\mu\nu}=0$. It plays a similar role as does
the Lorentz gauge $\partial_\mu A^\mu=0$ in electrodynamics. In the
weak-field approximation, $g_{\mu\nu}=\eta_{\mu\nu} +h_{\mu\nu}$
with $\eta_{\mu\nu}$ the Minkowski metric and $|h_{\mu\nu}|\ll 1$,
the harmonic gauge becomes
\begin{equation}
\partial^\mu h^\rho_{~\mu} -\frac 12 \partial^\rho h^\mu_{~\mu}=0.
\label{g1}
\end{equation}
(In the linear approximation, indices are raised and lowered with
the Minkowski metric, Greek indices run from 0 to 3, Latin indices
run from 1 to 3, and repeated indices are summed over.) By Eq.
(\ref{g1}), the linearized Einstein equation,
\begin{equation}
\square h_{\mu\nu}-\partial_\mu
\partial_\rho h^\rho_{~\nu}-\partial_\nu\partial_\rho
h^\rho_{~\mu}+\partial_\mu\partial_\nu h^\rho_{~\rho}=- S_{\mu\nu} ,
\label{Eins}
\end{equation}
reduces to the familiar form with a retarded solution:
\begin{equation}
\square h_{\mu\nu}=- S_{\mu\nu}. \label{harm}
\end{equation}
Here $\square \equiv \vec \partial ^2 -\partial_t^2$, $S_{\mu\nu}
\equiv T_{\mu\nu}-\frac 12 \eta_{\mu\nu} T^\rho_{~\rho}$, and we put
$16\pi G=1$.

The harmonic gauge, however, like the Lorentz gauge $\partial_\mu
A^\mu=0$, does not fix the gauge completely, and is not sufficient
to specify the two physical polarizations of the gravitational wave.
For pure gravitational waves without matter source, it was found
that the gauge freedom can be completely removed by the
transverse-traceless (TT) gauge~\cite{MTW70}:
\begin{subequations}
\label{TT}
\begin{eqnarray}
h_{0\mu}&=&0,\label{TT1}
\\
\partial_i h^i_{~j}&=&0 ,\label{TT2}\\
h^i_{~i}&=&0 .\label{TT3}
\end{eqnarray}
\end{subequations}

Nonetheless, as Ref.~\cite{MTW70} clearly illustrates, {\em only
pure waves (and not more general solutions of the linearized field
equations with source) can be reduced to TT gauge}. This can be
easily seen by counting the number of constraints: General
relativity admits a four-fold freedom of coordinate transformation,
but the TT gauge has altogether eight independent constraints, thus
can be satisfied only for special cases such as a pure wave, but not
for general cases.~\cite{noteCount} In fact, somehow surprisingly,
even for the linearized gravity we cannot find in the literature a
fully satisfactory gauge condition. Sometimes, the gauge $\partial^i
h^\rho_{~i}=0$ is called the ``Coulomb gauge''. But just like that
$\partial^\mu h^\rho_{~\mu}=0$ does not play a similar role in
gravity as $\partial_\mu A^\mu=0$ does in electrodynamics,
$\partial^i h^\rho_{~i}=0$ in gravity is not as convenient as
$\partial _i A^i=0$ in electrodynamics.

Resembling the relation between the Lorentz gauge $\partial_\mu
A^\mu=0$ and the radiation gauge $\partial _i A^i=0$, we find that
by a simple generalization from the harmonic gauge,
$g^{\mu\nu}\Gamma^\rho_{\mu\nu}=0$, we can get the ``true''
radiation gauge for gravity: $g^{ij} \Gamma ^\rho _{ij} =0$, or in
linearized form:
\begin{equation}
\partial^i h^\rho_{~i} -\frac 12 \partial^\rho h^i_{~i}=0.
\label{g2}
\end{equation}

This contains exactly four constraints, and we can check that they
can always be imposed: If $h_{\mu\nu}$ does not satisfy Eq.
(\ref{g2}), we can find a gauge-transformed $h'_{\mu\nu}$ that does.
Under an infinitesimal coordinate transformation: $x^\mu\to
x'^\mu=x^\mu -\epsilon ^\mu(x)$, with $\epsilon ^\mu $ four
arbitrary infinitesimal functions, $h_{\mu\nu}$ undergoes a gauge
transformation,
\begin{equation}
h_{\mu\nu}\to h'_{\mu\nu}=h_{\mu\nu}+\partial_\mu \epsilon _\nu +
\partial_\nu \epsilon _\mu.\label{trans}
\end{equation}
The goal can be achieved by choosing $\epsilon ^\mu$ with
\begin{equation}
\vec \partial ^2 \epsilon ^\rho =-(\partial^i h^\rho_{~i} -\frac 12
\partial^\rho h^i_{~i}). \label{para}
\end{equation}

By defining $\mathbbm{h}_{\mu\nu}=h_{\mu\nu}-\frac 12 \eta_{\mu\nu}
h^\rho_{~\rho}$, the harmonic gauge can be put in a concise
Lorentz-like form $\partial^\mu \mathbbm{h}^\rho_{~\mu}=0$.
Similarly, if we define ${\mathfrak h}_{\mu\nu}=h_{\mu\nu}-\frac 12
\eta_{\mu\nu} h^k_{~k}$, Eq. (\ref{g2}) can be put into a concise
transverse condition
\begin{equation}
\partial^i {\mathfrak h}^\rho_{~i} =0.
\label{g2'}
\end{equation}
It must be noted, however, that $\partial^i \mathbbm{h}^\rho_{~i}=0$
does not lead to the desired gauge (\ref{g2}). This might partially
explain why the gauge (\ref{g2}) has eluded people's attention.

It can be seen in two ways that Eq. (\ref{g2}) removes all
non-physical degrees of freedom of the gravitational field:
\begin{enumerate}
\item[(i)]  Eq. (\ref{g2}) permits no more gauge freedom.

\item[(ii)]  Eq. (\ref{g2}) leads to the equation of motion by which only two
physical components propagate.
\end{enumerate}

Proof of (i): Under a gauge transformation in (\ref{trans}),
preservation of the gauge condition (\ref{g2}) requires
\begin{equation}
\partial^i (\partial^\rho \epsilon_i+\partial_i \epsilon ^\rho)
 -\frac 12 \partial^\rho (\partial _i\epsilon ^i +\partial _i
 \epsilon ^i)=\vec \partial ^2 \epsilon ^\rho=0.
\label{test}
\end{equation}
By the boundary condition that $\epsilon ^\rho\to 0$ at infinity
\cite{noteBound}, the solution is $\epsilon^\rho \equiv 0$. With
such boundary condition, the solution to Eq. (\ref{para}) is also
fixed:
\begin{equation}
 \epsilon ^\rho =-\frac 1{\vec \partial ^2}(\partial^i h^\rho_{~i}
-\frac 12
\partial^\rho h^i_{~i}).
\end{equation}
Namely, the gauge transformation that brings an arbitrary (weak)
field tensor to the desired gauge in Eq. (\ref{g2}) is unique.
Incidently, this suggests a gauge-invariant construction:
\begin{equation}
\hat h_{\mu \nu}=h_{\mu \nu} - \frac 1{\vec\partial ^2}
(\partial_\mu\partial^i h_{i\nu}+\partial_\nu\partial^i h_{\mu i}
-\partial_\mu\partial_\nu h^i_{~i}). \label{GI}
\end{equation}

Proof of (ii): Apply the gauge (\ref{g2}), a careful examination
reveals that the gravitational equations of motion and gravitational
radiation resemble exactly the form of electrodynamics in the
radiation gauge:
\begin{subequations}
\label{G}
\begin{eqnarray}
\vec \partial ^2 h_{0\mu}&=&-S_{0\mu}, \label{G1}
\\
\square h_{ij} +\partial_t(\partial _j h_{0i}+\partial_i
h_{0j})-\partial_i \partial_j h_{00} &=&-S_{ij}. \label{G2}
\end{eqnarray}
\end{subequations}

To appreciate these equations and cast them into more elucidating
forms,  let us recall the corresponding ones in electrodynamics. In
the radiation gauge $\partial_i A^i=0$, the Maxwell equations
$\partial_\mu F^{\mu\nu}=-j^\nu$ take the form
\begin{subequations}
\label{E}
\begin{eqnarray}
\vec \partial^2  A^0&=&-j^0, \label{E1}
\\
\square A^i -\partial_t \partial ^i A^0 &=& -j^i . \label{E2}
\end{eqnarray}
\end{subequations}

Solving $A^0$ by Eq. (\ref{E1}), and using the conservation
condition $\partial_\mu j^\mu=0$, Eq. (\ref{E2}) can be casted into
\begin{equation}
\square \vec A_\perp=-\vec j_\perp, \label{E3}
\end{equation}
where $\vec j_\perp =\vec j-\vec \partial \frac 1{\vec \partial ^2}
(\vec \partial\cdot \vec j)$ is the transverse part of the electric
current. Eq. (\ref{E3}) has a very clear physical meaning: It is the
transverse field $\vec A_\perp$ with two independent components that
propagate in electromagnetic radiation, with the transverse current
$\vec j_\perp$ as its source.

Remarkably, we find that Eq. (\ref{G2}) can be casted into a form
analogous to Eq. (\ref{E3}):
\begin{equation}
\square \hat h_{ij}=-\hat S_{ij}.\label{hath}
\end{equation}
Here we have put a hat on $h_{\mu\nu}$ to remind that it is in the
radiation gauge and satisfies $\partial^i \hat h^\rho_{~i} -\frac 12
\partial^\rho \hat h^i_{~i}=0$. [It is also consistent to take this
$\hat h_{\mu\nu}$ as the gauge-invariant quantity in Eq.
(\ref{GI})]. Like $\vec A_\perp$, $\hat h_{ij}$ has also only two
(six minus four) independent components. They are the true dynamical
components that propagate in gravitational radiation. And similar to
$\vec j_\perp$, the radiation source $\hat S_{ij}$ is given by
\begin{equation}
\hat S_{ij}=S_{ij}-\frac {1}{\vec \partial ^2} (\partial_i\partial_k
S^k_{~j}+\partial_j \partial_k S^k_{~i} -\partial_i\partial_j
S^k_{~k}). \label{hatS}
\end{equation}

The derivation of Eqs. (\ref{hath}) and (\ref{hatS}) is exactly
analogous to that in electrodynamics: Solving $h_{0\mu}$ by Eq.
(\ref{G1}), and using the conservation condition in the weak-field
approximation, $\partial_\mu T^\mu_{~\nu}=0$, which implies
$\partial_\mu S^\mu _{~\nu}=\frac 12 \partial_\nu S^\mu_{~\mu}$. As
a critical cross check, it can be verified by Eq. (\ref{hatS}) that
\begin{equation}
\partial _i \hat S^i_{~j} -\frac 12
\partial_j \hat S ^i_{~i} =0,
\label{S}
\end{equation}
in consistent with $\partial^i \hat h^\rho_{~i} -\frac 12
\partial^\rho \hat h^i_{~i}=0$ in Eq. (\ref{hath}). Similar to
$\mathfrak {h}_{\mu\nu}$, we can define ${\mathfrak
S}_{\mu\nu}=S_{\mu\nu}-\frac 12 \eta_{\mu\nu} S^k_{~k}$ (again, {\em
not} ${\mathfrak S}_{\mu\nu}=S_{\mu\nu}-\frac 12 \eta_{\mu\nu}
S^\rho_{~\rho}$), and put Eq. (\ref{S}) into a transverse condition
$\partial _i \hat {\mathfrak S}^i_{~j}=0$. Then, similar to Eq.
(\ref{E3}), Eq. (\ref{hatS}) can be (loosely) interpreted as that
the gravitational radiation is generated by the ``transverse'' part
of the energy-momentum tensor.

The constructions in Eqs. (\ref{GI}) and (\ref{hatS}) are exactly
the same. In two other papers of this serial study
\cite{Chen10,Chen11}, we show that such construction applies quite
generally to any symmetric tensor field, and can be derived by
seeking a unique decomposition of a tensor field into physical and
pure-gauge components, analogous to the recent physical
decomposition of the Abelian and non-Abelian gauge fields by Chen
and collaborators \cite{Chen09,Chen08}.

The ``true'' radiation gauge we find has interesting relation to,
but crucial advantage over, the harmonic and TT gauges. This can be
most clearly seen for the pure wave solution in the vacuum. In the
harmonic gauge, the equations of motion (\ref{harm}) set all ten
field components to propagate, while the gauge conditions only
reduce the number of independent components to six. Thus, to obtain
the two physical polarizations, four additional constraints have to
be imposed by hand. The TT gauge, on the other hand, can promptly
pick out the two physical polarizations, but at the price of
``brutally'' imposing eight gauge conditions (which exceed the
actual number of gauge freedom and do not apply in general). The
exact counterpart of the TT gauge in electrodynamics is
\begin{subequations}
\begin{eqnarray}
A^0&=&0\\
\vec \partial \cdot \vec A&=&0
\end{eqnarray}
\end{subequations}
which, again, picks out the two physical polarizations of a free
electromagnetic field by ``brutally'' imposing more constraints than
the actual number of gauge freedom, and cannot be imposed in
general.

In our true radiation gauge, however, four field components
($h_{0\mu}$) are determined instantaneously by the source and are
non-dynamic, just like $A^0$ in electrodynamics. Especially, in the
vacuum with $S_{\mu\nu}=0$, we have automatically $h_{0\mu}=-\frac
1{\vec \partial ^2} S_{0\mu}\equiv 0$ by the trivial boundary
condition that $h_{0\mu}$ vanish at infinity. Then, of the six field
components $h_{ij}$ that propagate, the four gauge conditions in Eq.
(\ref{g2}) precisely pick out the two physical ones.

A closer look can reveal that with $h_{0\mu}=0$, Eq. (\ref{g2}) can
reproduce Eqs. (\ref{TT2}) and (\ref{TT3}): Set $\rho=0$ in Eq.
(\ref{g2}), we have $\partial^0 h^i_{~i}=2\partial^i h^0_{~j}=0$,
thus the spatial trace $h^i_{~i}$ is static, and actually must be
identically zero by the wave equation in the vacuum, $\square
h_{ij}=0$, together with a trivial boundary condition. Then by
setting $\rho =j$ in Eq. (\ref{g2}) we get $\partial ^i
h^j_{~i}=\frac 12 \partial ^j h^i_{~i}=0$, which is just Eq.
(\ref{TT2}). Namely, for a pure wave in our true radiation gauge, we
get the same eight constraints as the TT gauge by only four gauge
conditions plus four equations of motion. But unlike the TT gauge,
the ``true'' radiation gauge apply to general cases, not just pure
wave without source.

Very remarkably, the true radiation gauge also reveals how the TT
gauge can be adapted in the presence of source. From Eqs. (\ref{G}),
we can derive
\begin{subequations}
\label{TT'}
\begin{eqnarray}
h_{0\mu}&=&-\frac {1}{\vec \partial ^2} S_{0\mu},\label{TT1'}
\\
\partial_i h^i_{~j}&=&-\frac {1}{\vec \partial ^2} \partial_j T_{00} ,\label{TT2'}\\
h^i_{~i}&=&-2\frac {1}{\vec \partial ^2} T_{00} .\label{TT3'}
\end{eqnarray}
\end{subequations}
Namely, in the presence of source, $h_{0\mu}$, $h^i_{~i}$, and
$\partial_i h^i_{~j}$ cannot be all set to zero, but they can indeed
be chosen (in our true radiation gauge) to be non-dynamical. Since
matter source appear in Eq. (\ref{TT'}), we see that the TT
``gauge'' is rather a combination of field equations and real gauge
conditions. These real gauge conditions and (gauge-invariant) field
equations can be extracted from Eq. (\ref{TT'}) as follows: Act on
Eq. (\ref{TT3'}) with $\partial_j$, and compare the result with Eq.
(\ref{TT2'}), we get $\partial_ih^i_{~j}=\frac 12
\partial_j h^i_{~i}$. Then, act on Eq. (\ref{TT3'}) with $\partial_0$,
act on Eq. (\ref{TT1'}) with $\partial^\mu$, sum over $\mu$ from $1$
to $3$, and use the conservation relation $\partial^i S_{i0} =
\partial_0 T_{00}$, we get $\partial_i h^i_{~0}=\frac 12 \partial_0
h^i_{~i}$. Thus, (not surprisingly,) we get the full set of the real
gauge conditions in Eq. (\ref{g2}) which are solely constrains on
$h_{\mu\nu}$ and independent of the field equations. Then, with
these gauge conditions, Eq.~(\ref{TT1'}) can recover its
gauge-invariant form in Eq.~(\ref{Eins}). Thus we see clearly that
the ``adapted TT ``gauge'' in Eq. (\ref{TT'}) is indeed a mixture of
really independent gauge conditions with original (gauge-invariant)
Einstein equations. In retrospect, the original TT gauge in Eq.
(\ref{TT}), which is obtained automatically from Eq. (\ref{TT'}) by
setting source term to zero, is also of a mixed nature. Actually, by
repeating the above procedure, we can extract from Eq. (\ref{TT})
the real gauge conditions in Eq. (\ref{g2}), together with
gauge-invariant source-free equations for $h_{0\mu}$, which are just
the ($0\mu$)-components of Eq. (\ref{Eins}) without source.

It is worthwhile to discuss a ``minimum-TT'' gauge, which is defined
by combining Eqs. (\ref{TT2}) and (\ref{TT3}), and discarding Eq.
(\ref{TT1}). In the celebrated canonical formulation of gravity by
Arnowitt, Deser, and Misner (ADM) \cite{ADM}, this minimum-TT gauge
is regarded as the radiation gauge for gravity, since in this gauge
$h_{ij}$ has only TT components, which in the ADM formulation are
the dynamical variables of the gravitational field. We point out
that this minimum-TT gauge does not work for general cases either.
First, by setting no constraint on coordinate transformation
involving only $x^0$, it does not fix the gauge completely. Second,
it sets four constraints on purely ``spatial'' transformations,
which admits however only a three-fold gauge freedom. Thus, like the
complete TT gauge, the minimum-TT gauge can be imposed only for
special cases such as a pure wave, but not general cases. In other
words, it is not always possible to transform a general symmetric
tensor into purely TT form. Certainly, the combination of
Eqs.(\ref{TT2'}) and (\ref{TT3'}) can be regarded as an adaption of
the minimum-TT gauge in the presence of source. But as we just
explained, they are rather a combination of true gauge conditions
with field equations.

In electrodynamics, the Lorentz gauge $\vec \partial \cdot \vec
A+\partial_t A^0=0$ and radiation gauge $\vec \partial \cdot \vec
A=0$ coincide for a static configuration. This property, however, is
no longer true in gravitation. This can be seen by comparing Eqs.
(\ref{g1}) and (\ref{g2}). They differ by a term $\partial^0
h^\rho_{~0} -\frac 12 \partial^\rho h^0_{~0}$, in which
$\partial^\rho h^0_{~0}$ can be non-zero even for a static case. We
can see the crucial difference more directly by looking at the
equations of motion. For a static configuration, in the harmonic
gauge we have $\vec \partial ^2 h_{\mu\nu}=-S_{\mu\nu}$, while in
the radiation gauge we have $\vec \partial ^2 h_{0\mu }=-S_{0\mu}$
but
\begin{equation}
\vec \partial ^2 h_{ij} -\partial_i \partial_j h_{00} =-S_{ij},
\label{Gs}
\end{equation}
which differs significantly from that in the harmonic gauge. In the
following, we present explicitly an interesting example for the
simplest spherically symmetric solutions. We consider a weak source
with a constant mass density $\rho_0$ and negligible pressure:
$T_{\mu\nu}={\rm diag}\{\rho_0, 0,0,0\}$, and $S_{\mu\nu}=\frac 12
{\rm diag}\{\rho_0, \rho_0,\rho_0,\rho_0\}$, distributed within a
radius $R$ with total mass $M$. In the harmonic gauge, we find the
external solution (with $G$ displayed for clarity)
\begin{equation}
h_{\mu\nu}=\delta _{\mu\nu} \frac {2GM}r,
\end{equation}
\begin{equation}
ds^2=-(1-\frac{2GM}r) dt^2+(1+\frac{2GM}r)d\vec x^2, \label{Sch1}
\end{equation}
This is just the first-order approximation to the familiar
Schwarzschild solution. In the radiation gauge, we find for the
external area the same $h_{0\mu}$ but a distinct $h_{ij}$:
\begin{equation}
h_{0\mu}=\delta _{0\mu} \frac {2GM}r,~ h_{ij}=(3 \delta _{ij}- \frac
{x_i x_j}{r^2}) \frac {GM}r,
\end{equation}
\begin{equation}
ds^2=-(1-\frac{2GM}r dt^2)+(1+\frac{3GM}r)d\vec x^2-\frac{GM}r
\frac{(\vec x\cdot d\vec x)^2}{r^2}. \label{Sch2}
\end{equation}
This form appears rather peculiar and has never been discussed
before. We will return to its use shortly below.

Besides its own importance as a novel and complete gauge condition
for exploring the dynamical structure of gravitation~\cite{note},
the major relevance of the ``true'' radiation gauge is for
clarifying the long-standing problem of gravitational energy {\em
distribution}. With the increasing opportunity of detecting a
gravitational wave and using it as a novel probe into the university
\cite{GW}, one must now take seriously a long suspended
controversial problem, namely, the gauge dependence of energy flow
in gravitational radiation. The key obstacle to assigning an
unambiguous energy to the gravitational field is the fact that the
metric field $g_{\mu\nu}$ contains spurious gravitational effect
associated with coordinate choice. Now that we have found a complete
gauge condition which can remove all non-physical degrees of freedom
of the gravitational field, the gravitational energy can be safely
calculated in this radiation gauge. E.g., it is Eq. (\ref{Sch2})
rather than Eq. (\ref{Sch1}) that provides a pertinent metric for
calculating the energy density of gravitational field ~\cite{Naka}.
More importantly, when discussing the distribution of energy flow in
gravitational radiation, one should always work in the ``true''
radiation gauge (\ref{g2}), or equivalently in the TT gauge if one
concerns about the radiation zone far away from the source. The
latter approach was adopted in Ref.~\cite{Hart03}. And we should
give (a somewhat academic) reminder that if one concerns about the
energy flow near the radiation source, the primitive TT gauge in Eq.
(\ref{TT}) cannot be imposed, and only the radiative gauge
(\ref{g2}), or equivalently the ``adapted TT gauge'' in Eq.
(\ref{TT'}), is appropriate. Another but necessary reminder:
Radiative solutions in the harmonic gauge are the easiest to obtain,
but must not be directly employed to compute the angular
distribution of energy flow~\cite{Wein72}, since $g_{\mu\nu}$ in
this gauge contain non-physical components which represent a
spurious gravitational effect.

{\it Discussion.---}In this paper we have focused on the weak-field
regime and infinitesimal gauge transformations. We now briefly
discuss the general cases. We have seen that Eq. (\ref{g2}) appear
to be the unique choice for linearized gravity. The extension of Eq.
(\ref{g2}) to general cases, however, is not unique. E.g.,
$g^{ij}\Gamma^\rho_{ij}=0$, $\eta ^{ij}\Gamma^\rho_{ij}=0$, and
$g^{i\mu}\Gamma^\rho_{i\mu}=0$ all reduce to Eq. (\ref{g2}) in the
linear approximation. It is not easy to tell which choice is better.
Beyond the weak-field approximation, the gauge transformation of the
gravitational field shows a non-Abelian character:
\begin{equation}
g'_{\mu\nu}=g_{\mu\nu}+(\partial_\lambda g_{\mu\nu}) \epsilon
^\lambda +g_{\mu\lambda }\partial_\nu  \epsilon ^\lambda
+g_{\nu\lambda }\partial_\mu  \epsilon ^\lambda . \label{transg}
\end{equation}
As the lesson learned from non-Abelian gauge theories, for large
field amplitude and large gauge transformation, it should not be
expected that the gauge can be completely fixed by simple algebraic
and differential constraints in the non-linear theory. The situation
for gravity would by no means be simpler, and we leave this highly
complicated and non-trivial issue to future studies. For the moment,
the only justification we have for preferring
$g^{ij}\Gamma^\rho_{ij}=0$ is similar to that in defining the
harmonic gauge $g^{\mu\nu}\Gamma^\rho_{\mu\nu}=0$. In the harmonic
gauge, the coordinate-invariant d'Alembertian $g^{\mu\nu} D_\mu
D_\nu$ (where $D_\mu$ is the covariant derivative) reduces to the
ordinary form:
\begin{equation}
g^{\mu\nu} D_\mu D_\nu \phi =g^{\mu\nu} \partial_\mu \partial_\nu
\phi -g^{\mu\nu} \Gamma^\sigma  _{\mu\nu} \partial_\sigma \phi=
g^{\mu\nu} \partial_\mu \partial_\nu \phi,
\end{equation}
so that the coordinates $x^\rho$ are harmonic, in the sense of
obeying
\begin{equation}
g^{\mu\nu} D_\mu D_\nu x^\rho =g^{\mu\nu}
\partial_\mu \partial_\nu x^\rho =0.
\end{equation}

In our radiation gauge $g^{ij}\Gamma^\rho_{ij}=0$, it is the
invariant Laplacian $g^{ij} D_i D_j$ that reduces to the ordinary
form:
\begin{equation}
g^{ij} D_i D_j \phi =g^{ij} \partial_i \partial_j \phi -g^{ij}
\Gamma^\sigma  _{ij} \partial_\sigma \phi= g^{ij} \partial_i
\partial_j \phi.
\end{equation}
Now, the coordinates $x^\rho$ obey the Laplace equation
\begin{equation}
g^{ij} D_i D_j x^\rho =g^{ij}
\partial_i \partial_j x^\rho =0.
\end{equation}
Due to the instantaneous feature of the Laplace equation, this can
be interpreted as that the coordinate $x^\rho$ are non-dynamic and
do not propagate, thus justifies our choice
$g^{ij}\Gamma^\rho_{ij}=0$ as the true radiation gauge for the
gravitational field. Namely, in this gauge only the gravitational
degrees of freedom propagate and the  ``coordinate wave'' is absent.

We are grateful to the referee for very helpful comments. This work
is supported by the China NSF Grants 10875082 and 11035003. XSC is
also supported by the NCET Program of the China Education
Department.


\begin{thebibliography}{99}
\bibitem{noteCount} Note that a gauge may not work if its number of
constraints does not exceed the actual degrees of freedom, but is
doomed to fail if it contains more independent constraints than the
actual degrees of freedom.
\bibitem{noteBound} In this paper we use frequently such trivial
boundary condition at infinity. Note that this does not exclude the
discussion of gravitational wave or radiating system, which can be
made spatially finite by considering a wavepacket or by letting the
radiation occur in a given period. The real case that necessitates a
non-trivial boundary condition and requires special care is the
universe, which we leave for future studies.
\bibitem{note} We conjecture that the canonical structure and
quantization of gravity in the ``true'' radiation gauge would be
particularly illuminating.
\end{thebibliography}
\end{document}